\documentclass[aps,prc,bibnotes,superscriptaddress,nofootinbib,showpacs,floatfix,twocolumn]{revtex4-1}
\usepackage[colorlinks,linkcolor=blue,citecolor=blue,filecolor=black,urlcolor=blue]{hyperref}
\usepackage{graphicx} 
\usepackage{soul} 
\usepackage{xspace} 
\usepackage{amsmath}
\usepackage{harpoon}
\usepackage{appendix}
\usepackage{color}
\RequirePackage{lineno} 



\makeatletter 
\@addtoreset{equation}{section}
\makeatother  

\newcommand{\sqrtsnn}{\mbox{$\sqrt{s_{\mathrm{NN}}}$}}

\newcommand{\pT} {p_{\mathrm{T}}}
\newcommand{\lr}[1]{\left\langle #1\right\rangle}
\newcommand{\lrb}[1]{\left[ #1\right]}

\begin{document}
\title{Imaging nuclei by smashing them at high energies: how are their shapes revealed after destruction?}
\newcommand{\sbu}{Department of Chemistry, Stony Brook University, Stony Brook, NY 11794, USA}
\newcommand{\bnl}{Physics Department, Brookhaven National Laboratory, Upton, NY 11976, USA}
\author{Jiangyong Jia}\email[Correspond to\ ]{jiangyong.jia@stonybrook.edu}\affiliation{\sbu}\affiliation{\bnl}
\begin{abstract}
High-energy nuclear collisions have recently emerged as a promising ``imaging-by-smashing'' approach to reveal the intrinsic shapes of atomic nuclei. Here, I outline a conceptual framework for this technique, explaining how nuclear shapes are encoded during quark-gluon plasma formation and evolution, and how they can be decoded from final-state particle distributions. I highlight the method's potential to advance our understanding of both nuclear structure and quark-gluon plasma physics.
\end{abstract}
\maketitle
Imaging techniques are fundamental to scientific discovery, enabling us to visualize and understand structures across vastly different scales. These methods employ external probes, often photons, to interact with samples and glean structural information. The sample's structural information is encoded in the scattered probe. In X-ray crystallography, for example, a molecule's three-dimensional organization is deduced from momentum-space scattering patterns. The imaging process takes place as the probe traverses the sample, and because any structural disintegration happens only after scattering, the captured information remains intact.

A radically different approach has emerged in nuclear physics: rather than using external probes, atomic nuclei are collided at ultra-high energies to create a hot, dense state of matter called the quark-gluon plasma (QGP). Remarkably, the geometric structure of the colliding nuclei determines the QGP's initial conditions and influences its subsequent evolution. This process ultimately imprints specific patterns on the momentum distributions of thousands of final-state particles. By carefully ``rewinding'' this evolution, one can reconstruct an effective image of the original nuclei.

The STAR Collaboration demonstrated this concept by extracting the shape of $^{238}$U from $^{238}$U+$^{238}$U collision data that are consistent with low-energy measurements. Despite completely destroying the nuclei, their structural information is preserved with sufficient detail to allow reconstruction~\cite{STAR:2024wgy}. This breakthrough raises fundamental questions: How are nuclear shapes encoded in such violent collisions? How can they be deduced from final-state particles? What are the broader implications for nuclear physics?

{\bf Traditional methods for nuclear shapes.}
Atomic nuclei are bound systems of protons and neutrons (nucleons) held together by the strong nuclear force. While often depicted as spheres, nuclei can adopt prolate, oblate, or even pear-like configurations -- shapes that reflect their underlying many-body wavefunctions. These wavefunctions govern the positions and momenta of nucleons. The non-perturbative nature of the strong force makes theoretical predictions of these shapes challenging and experimental input crucial. 

Unlike molecules, nuclei cannot be oriented in a crystalline lattice for direct coherent imaging. Instead, low-energy techniques such as electron scattering, laser spectroscopy, and Coulomb excitation (Coulex) have been used to infer nuclear shapes~\cite{Yang:2022wbl}. Electron scattering probes nuclei one at a time. This provides an orientation-averaged image where nuclear deformation appears as modifications to charge distributions. Laser spectroscopy probes nuclear shapes through measurements of hyperfine atomic transitions. Coulex uses low-energy ions to excite nuclei into rotational and vibrational states. Nuclear shapes are then deduced by detecting gamma rays emitted during relaxation and comparing the measured emission rates to theoretical models.

However, these methods rely on electromagnetic interactions and thus cannot directly probe neutron distributions. Moreover, their relatively long timescales yield time-averaged, blurred representations of nuclear shapes, akin to long-exposure photographs. This temporal averaging obscures rapid fluctuations and dynamical variations in nucleon distributions that affect nuclear shapes.

\begin{figure*}[htbp] \centering
\includegraphics[width=0.9\linewidth]{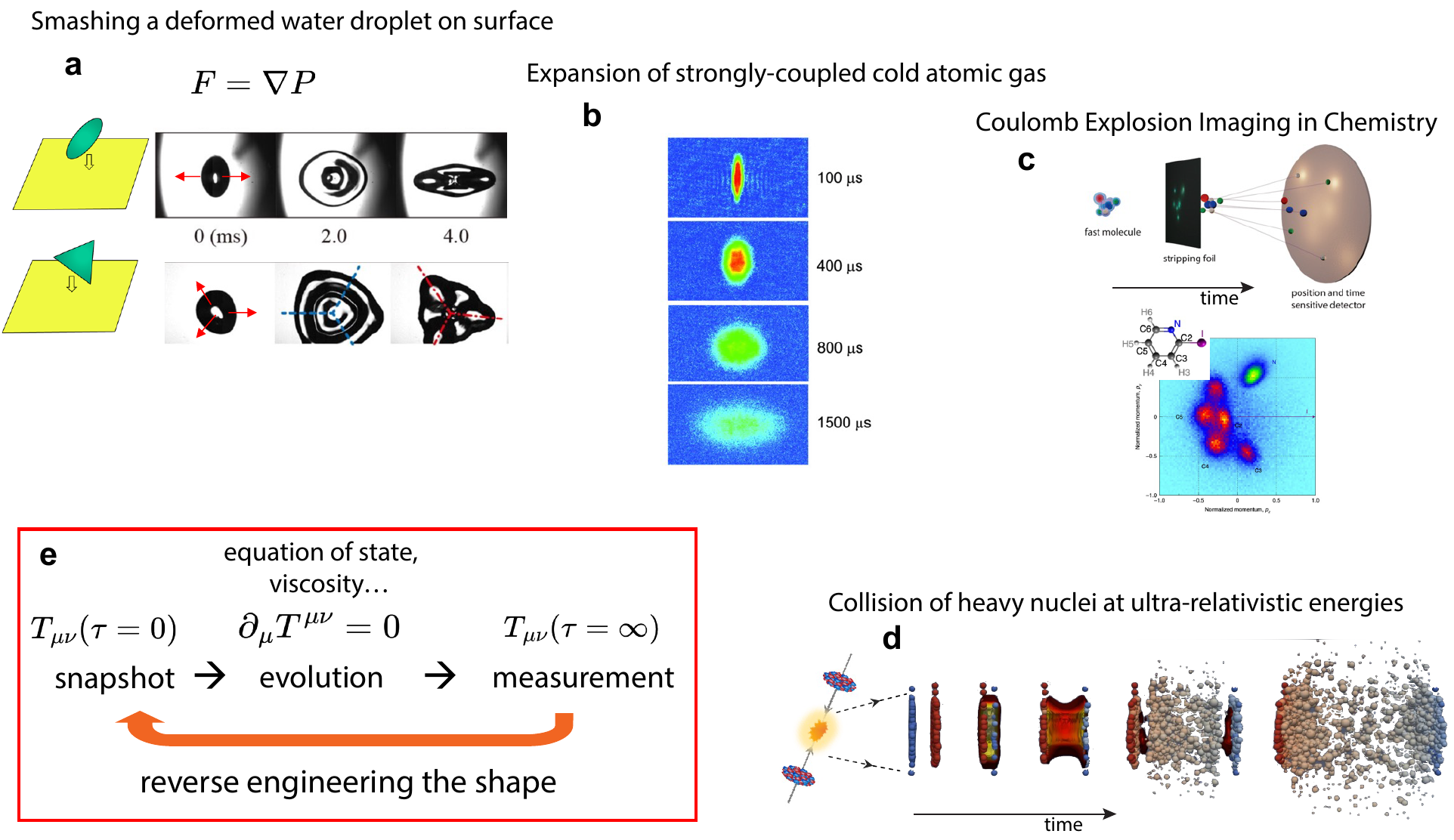}
\caption{\label{fig:1} {\bf Connections between initial and final state in various ``smashing'' experiments:} 
(\textbf{a}) Deformed water droplet colliding with a hydrophobic surface, producing an expansion pattern that inverts the initial shape asymmetry~\cite{Yun:2017};
(\textbf{b}) Expansion of a strongly-coupled Fermi gas released from an optical trap~\cite{OHara:2002pqs}, whose geometry leaves imprints in the subsequent dynamics of the gas due to the ultra-fast switch-off of the confining potential;
(\textbf{c}) Coulomb-explosion imaging of a small molecule stripped of electrons, in which nuclear positions are inferred by reversing the repulsive expansion~\cite{cei,ceinature};
(\textbf{d}) Pressure-driven expansion of the quark-gluon plasma produced in high-energy nuclear collisions~\cite{madai}.
In each case, the final state can be reverse-engineered to extract the initial condition, provided the expansion dynamics, represented by the system's energy-momentum tensor $T_{\mu\nu}$, are sufficiently well understood (\textbf{e}). Note that panel-\textbf{e} draws an analogy in their response patterns rather than a direct comparison of the physics.}
\end{figure*}

{\bf Imaging-by-smashing at high energy.} 
High-energy nuclear collisions offer a fundamentally different imaging paradigm (Fig.~\ref{fig:1}d). The key innovation lies in the extremely short interaction timescales, set by the crossing time of the two colliding nuclei $\tau=2 R_0 / \Gamma < 0.1 \mathrm{fm} / \mathrm{c}$, where $R_0$ is the nuclear radius and $\Gamma> 100$ is the Lorentz contraction factor. This timescale is so brief that it effectively provides an instantaneous ``snapshot'' of the nuclear configuration.

From a quantum mechanical perspective, a nucleus in its isolated ground state exists in a superposition of all possible orientations, making its intrinsic shape unobservable~\cite{Verney:2025efj}. However, high-energy heavy-ion collisions represent strongly interacting, dynamical systems where nuclear ground states are not eigenstates of the combined system. The extremely rapid collision process acts as a quantum measurement, effectively collapsing the nuclear many-body wavefunction and projecting nucleons into specific positions, from which nuclear shape and orientation can be defined.

Initially, the nucleus's shape directly influences the conditions under which the QGP forms. As the plasma expands hydrodynamically, information about the original nuclear shape transforms and becomes encoded in the momentum distribution of thousands of final-state particles. By analyzing these distributions and reverse-engineering the expansion process event by event, nuclear shapes can be reconstructed. The precision of this method hinges on how well the initial conditions, equation of state, and transport properties of the QGP are understood. These aspects have been the focus of the heavy-ion community for decades and are now believed to be well-constrained~\cite{Busza:2018rrf}.

This imaging-by-smashing principle appears across diverse physical systems and length scales, as shown in Fig.~\ref{fig:1}a--c. For example, deformed water droplets colliding with hydrophobic surfaces undergo pressure-driven expansions that invert their shape asymmetries~\cite{Yun:2017}. Strongly-coupled Fermi gases released from optical traps exhibit anisotropies in their final states that reflect initial geometries~\cite{OHara:2002pqs}. Chemistry provides another parallel through Coulomb explosion imaging (CEI), where removing electrons from a molecule using a laser or thin foil induces nuclear repulsion that can be ``unwound'' to reveal the molecule's spatial configuration~\cite{cei,ceinature}. These examples share three essential features (Fig.~\ref{fig:1}e): an initial configuration to be imaged, collective expansion governed by well-defined evolution equations, and final state detection. Successful imaging requires an adequate understanding of the expansion dynamics to reverse-engineer the original structure.

Despite its seemingly extreme nature, high-energy nuclear imaging may offer distinct advantages. First, the collision probes the {\it ground state nuclear many-body wavefunction in position space}, information not readily accessible through more conventional means. The colliding nuclei and their Coulomb fields are Lorentz-contracted, ensuring that the nuclei are causally disconnected before the collision. This causal disconnection prevents pre-collision excitation effects that could contaminate the measurement. While nuclei could be Coulomb excited when they pass by each other, most isomeric states are very short-lived (typically $<10^{-12}$ s) and cannot persist in the circulating beam long enough to participate in the subsequent collisions~\footnote{Some isomers have exceptionally long lifetime, such as 7.7s 11/2$^-$ state of $^{197}$Au and 8.9 days 11/2$^-$ state of $^{129}$Xe, which can accumulate in the beam. However, no estimates are currently available on their production rates in high-energy accelerators.}

Second, the subsequent expansion is well described by classical hydrodynamics, an effective theory valid for systems with densely populated degrees of freedom (DOF), irrespective of their sizes. This provides a robust theoretical framework for interpreting the measurements. Third, high-energy collisions produce far more final-state particles per event than any conventional experiments, allowing for event-by-event reconstructions with unprecedented statistical precision. In this sense, the process's apparent destructiveness becomes an asset.

{\bf Energy dependence of the nuclear image.}
The observed nuclear image, and its effective many-body wavefunction, depends critically on the collision energy $\sqrtsnn$, which determines the effective ``shutter speed'' of the imaging process. At low energies, where the shutter speed is slow, collective rotational and vibrational DOF dominate the nucleus's apparent shape. For instance, the rotational DOF of a deformed heavy nucleus, with excitation energies around 0.1 MeV, corresponds to timescales of $\tau \sim 10^3-10^4$ fm/$c$. As collision energy increases, faster modes such as nucleon clustering and short-range correlations come into play. At still higher energies, the resonance structure of nucleons emerges, eventually giving way to the subnucleonic quarks and gluons. The imaging process effectively captures all DOF slower than the nuclear crossing timescale. As a result, varying $\sqrtsnn$ provides a natural way to study the evolution of the nuclear wavefunction across energy scales.

\begin{figure*}[htbp] \centering
\includegraphics[width=0.7\linewidth]{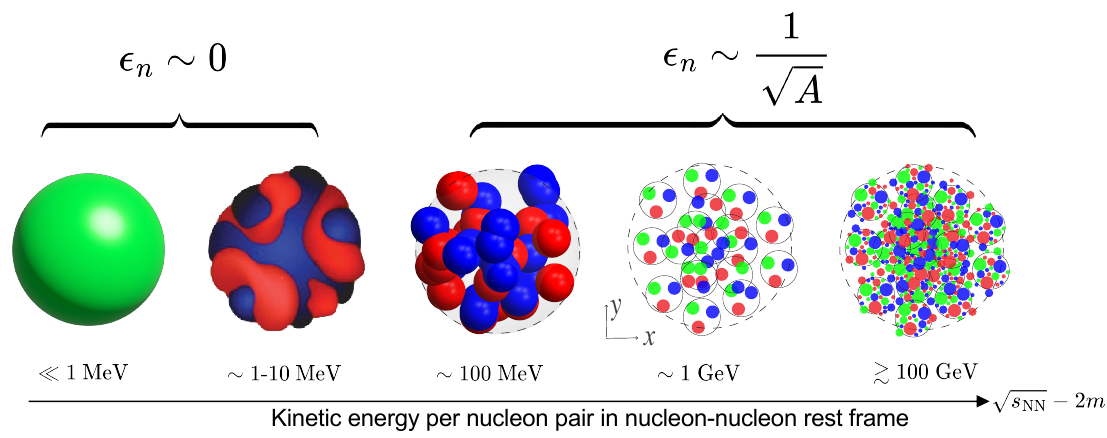}
\caption{\label{fig:2} {\bf Energy dependence of nuclear structure.} Different degrees of freedom become relevant at different energies, affecting the apparent shape of the nucleus. Due to quantum fluctuations at nucleon and subnucleonic level, even a nominally spherical nucleus such as $^{208}$Pb exhibits a deformed nucleon distribution in the transverse plane at high energies.}
\end{figure*}

While electron-nucleus scattering probes the one-body distribution of the scattering centers -- protons at low energy and quarks and gluons at high energy -- much less is known about the nuclear many-body distributions across energy scales. High-energy heavy-ion collisions help fill this gap. Measurements of anisotropic flow coefficients reveal quantum fluctuations associated with the finite number of scattering centers~\cite{Heinz:2013wva}. These quantum fluctuations induce deformation in the transverse ($xy$) nucleon distribution, with eccentricities scaling roughly as $1/\sqrt{A}$, where $A$ is the mass number~\footnote{The eccentricities are mostly determined by the positions of nucleons, as most quarks and gluons are confined within them and nuclear modifications are modest.}. This fluctuation-driven deformation produces non-zero flow even in head-on collisions~\cite{CMS:2013bza}, despite the nuclear overlap region being isotropic on average (see Fig.~\ref{fig:2}). 

Based on these considerations, nuclear images at high energy thus contain two distinct components. The first reflects slow, global shape modes, such as rotational and vibrational, independent of collision energy. The second arises from quantum fluctuations at the nucleon and subnucleonic scales, which vary with $\sqrtsnn$. A robust imaging method is essential for disentangling and studying each contribution separately. By comparing results from RHIC and the LHC, one can investigate the energy dependence of these two components in detail.

\begin{figure*}[htbp] \centering
\includegraphics[width=0.7\linewidth]{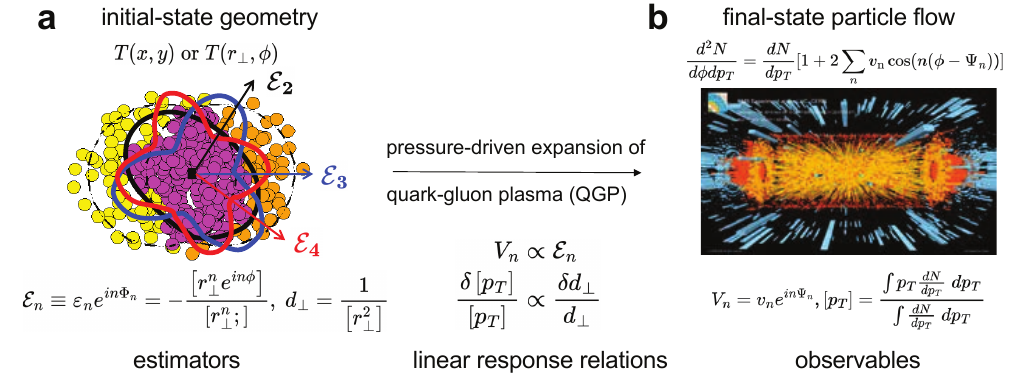}
\caption{\label{fig:3} {\bf Relation between initial- and final-state of high-energy nuclear collisions.} The features of collision geometry are characterized by its shape and size parameters, $\mathcal{E}_n$ and $d_{\perp}$ (${\bf a}$). They are linearly related to observables that describe the transverse momentum $\pT$ spectra in each event, i.e., the anisotropic flow coefficients $V_n$ and the average transverse momentum $[\pT]$ (${\bf b}$). The event-to-event variations of these initial- and final-state quantities are linearly related.}
\end{figure*}

{\bf Reverse-engineering nuclear shape.} Having established the conceptual foundation, we now turn to the practical implementation of nuclear shape reconstruction. A nucleus with quadrupole deformation can be described by a surface function in terms of the polar angle $\theta$ and azimuthal angle $\phi$,
\begin{align}\label{eq:1}
R(\theta,\phi) = R_0(1+\beta_2 (\cos \gamma Y_{2,0}+ \sin\gamma Y_{2,2}))\;,
\end{align}
where  $Y_{l,m}(\theta,\phi)$ are spherical harmonics in real bases. The parameters $\beta_2$ and $\gamma$ define the quadrupole deformation and triaxiality, respectively. The $\gamma$ controls the ratios of principal radii. As $\gamma$ varies from $0^{\circ}$ to $60^{\circ}$, the nucleus transitions from prolate ($\gamma=0^{\circ}$) to oblate ($\gamma=60^{\circ}$), with intermediate values corresponding to triaxial shapes. Higher-order deformations, such as octupole $\beta_3$ or hexadecapole $\beta_4$, can also be included but are typically smaller than $\beta_2$. When projected onto the $xy$-plane, these deformation components give rise to elliptic, triangular, or quadrangular initial geometries of the QGP~\cite{Jia:2021tzt}.

Imaging nuclear shape at high energy involves a three-step process. First, the initial conditions of the collision are reconstructed from final-state observables using hydrodynamic models that link measured particle distributions to the geometry of the QGP. Second, these initial conditions are mapped back to the intrinsic nuclear shape. Finally, by comparing two isobar-like collision systems -- nuclei with similar mass but differing structural properties -- global nuclear shape effects can be distinguished from fluctuations on nucleonic and subnucleonic scales.

\begin{figure*}[htbp] \centering
\includegraphics[width=0.92\linewidth]{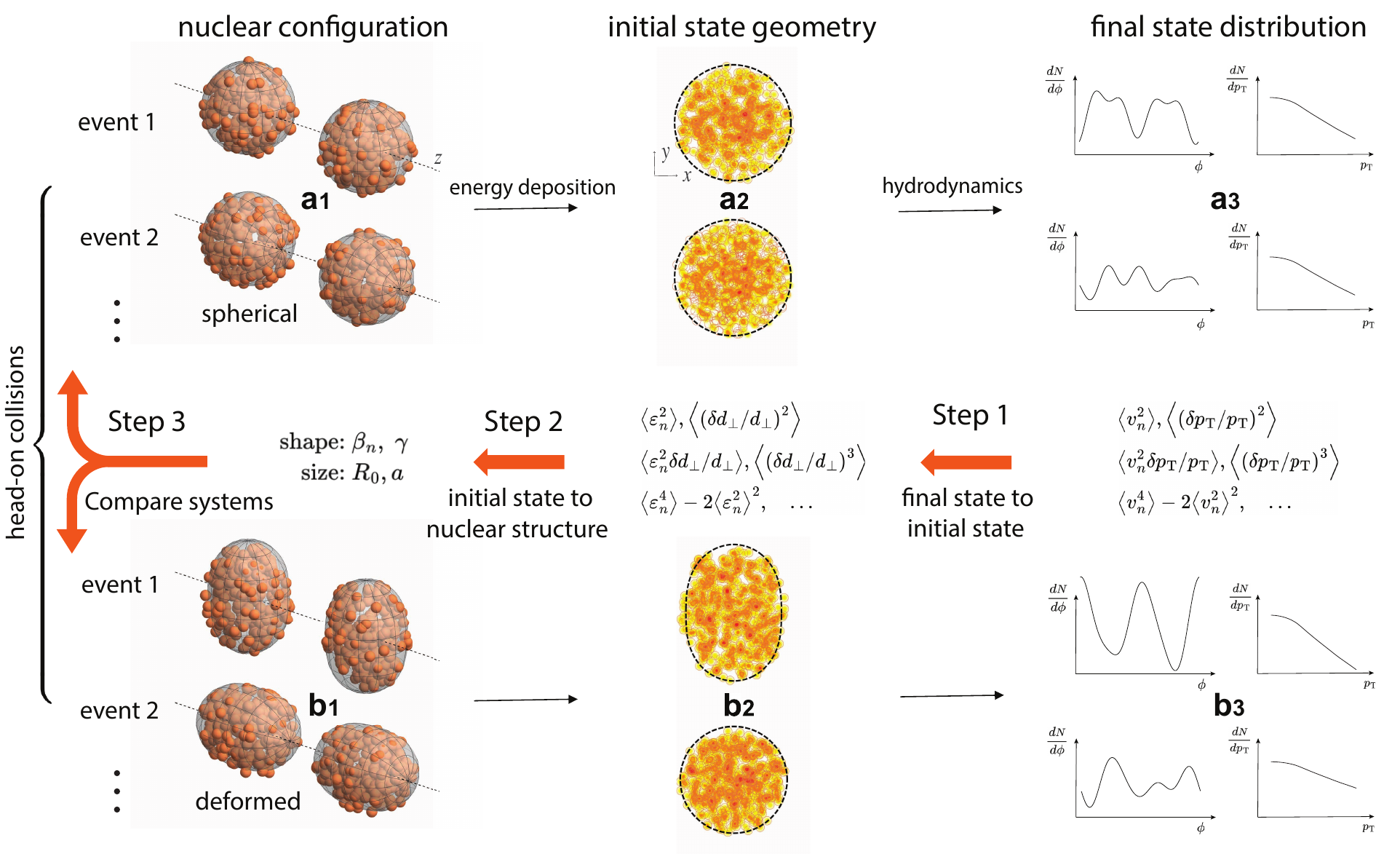}
\caption{\label{fig:4} {\bf Three steps in the imaging-by-smashing method.} 
Illustrated here are collisions of spherical nuclei (${\bf a1}$--${\bf a3}$) and prolate-deformed nuclei  with $\beta_2=0.28$ (${\bf b1}$--${\bf b3}$) for two representative collision events: (${\bf a1}$ and ${\bf b1}$) initial configurations of the colliding nuclei, (${\bf a2}$ and ${\bf b2}$) initial geometry in the transverse plane, and (${\bf a3}$ and ${\bf b3}$) final-state distribution of particles in azimuthal angle and $\pT$. The event-to-event variation in the initial-state geometry and final-state distributions is quantified by the moments involving $\varepsilon_n$ and $d_{\perp}$ of the initial state and $v_n$ and $[\pT]$ of the final state, as indicated in the middle row. Parameter $a$ represents the nuclear surface diffusivity. The nucleon positions are simulated in a Monte-Carlo Glauber model with $R_0=6.81$~fm, $a=0.55$~fm, and $A=238$. Imaging involves reconstructing the initial geometry from final-state particles (Step 1), relating it to the nuclear shape, which is affected by both global deformation and quantum fluctuations (Step-2), and comparing different collision systems to isolate the global deformation (Step-3).}
\end{figure*}

\ul{\textit{Step1: From the final state to the initial condition}}.
The QGP's initial geometry in a single event is characterized by its energy density distribution $T(x,y)$ (Fig.~\ref{fig:3}\textbf{a}), where $x$ and $y$ are distribution's principal axes relative to its center-of-mass. Key parameters include the total energy $E= \int T(x,y) dxdy$, the elliptic and triangular eccentricities $\varepsilon_2$ and $\varepsilon_3$, and inverse of mean-square area $d_{\perp}$:
\begin{align}\nonumber
d_{\perp} &= 1/\sqrt{\lrb{x^2}\lrb{y^2}}\;, \\\label{eq:2}
\mathcal{E}_n\equiv \varepsilon_n e^{i n\Phi_n} &= - \lrb{(x+iy)^n}/\lrb{|(x+iy)|^n},
\end{align}
with ``$\lrb{..}$'' indicates averaging weighted by $T(x,y)$, and $\Phi_n$ represents the orientation of the $n^{\rm th}$ eccentricity. 

The total energy $E$ influences the number of charged particles. Meanwhile, the initial eccentricites $\varepsilon_n$ drive anisotropic flow, described by \mbox{$dN/d\phi~\propto~1+2\sum_nv_n\cos(n(\phi-\Psi_n))$}, where $v_n$ (with phase $\Psi_n$) define the elliptic ($v_2$) and triangular ($v_3$) flow coefficients (Fig.~\ref{fig:3}\textbf{b}). Additionally, $d_{\perp}$ governs the radial expansion or ``radial flow'', affecting the average transverse momentum $[\pT]$ of final-state particles. Hydrodynamic modeling suggests approximately linear relationships between these initial- and final-state observables. Specifically, $v_n\propto \varepsilon_n$ and $\delta \pT/\pT \propto \delta d_{\perp}/d_{\perp}$~\cite{Niemi:2012aj,Bozek:2012fw}~\footnote{Here, $\frac{\delta \pT}{\pT} = \frac{[\pT]-\lr{[\pT]}}{\lr{[\pT]}}$ and $\frac{\delta d_{\perp}}{d_{\perp}} = \frac{d_{\perp}-\lr{d_{\perp}}}{\lr{d_{\perp}}}$ denote event-by-event fluctuations, and ``$\lr{..}$'' indicates average over events.}. The proportionality constants (response coefficients) are strongly impacted by QGP properties.

Event-by-event variations in nuclear geometry arise from both global deformation and quantum fluctuations. These variations are quantified by moments of initial- and final-state observables. These moments are connected by the linear response relations shown in Fig.~\ref{fig:3}. Our understanding of the initial conditions, evolution, and properties of the QGP is mainly derived from the study of these relations~\cite{Heinz:2013wva,Busza:2018rrf}.  Many such moments can be used, as indicated in the middle row of Fig.~\ref{fig:4}. The STAR collaboration specifically focused on three: $\lr{v_2^2}$, $\lr{(\delta \pT)^2}$, and $\lr{v_2^2\delta \pT}$~\cite{STAR:2024wgy}.

Extracting QGP properties, such as the equation of state and transport coefficients, has long been a major goal in heavy-ion physics. The state-of-the-art approach employs Bayesian analyses to simultaneously constrain these QGP parameters and the initial conditions \cite{Bernhard:2019bmu}. However, uncertainties in the initial conditions still limit the precision of these extractions -- a limitation that can be reduced by leveraging the collision of species with well-understood shapes \cite{Jia:2022ozr}.

\ul{\textit{Step2: from the initial conditions to nuclear shape}}. 
The QGP's initial condition is closely tied to nucleon distributions in colliding nuclei A and B, described by thickness functions $T_{\rm A,B}(x,y)$, which fluctuate from event to event. In the presence of global deformation, these functions also depend on the nuclei's orientations prior to collision (see Fig.~\ref{fig:4}\textbf{a1} and \ref{fig:4}\textbf{b1}). For head-on collisions, random orientations of prolate deformed nuclei lead to significant, anti-correlated fluctuations between $v_2$ and $[\pT]$~\cite{Giacalone:2019pca}. General considerations imply that such fluctuations follow simple parametric forms \cite{Jia:2021qyu}. Specifically, the three moments used by the STAR Collaboration are~\footnote{Shape fluctuations can be accounted for by replacing $\beta_2^2$ and $\beta_2^3\cos(3\gamma)$ by $\lr{\beta_2^2}$ and $\lr{\beta_2^3\cos(3\gamma)}$, respectively.}:
\begin{align}\nonumber
\lr{v_2^2}&= a_1+b_1\beta_2^2\;,\\\nonumber
\lr{(\delta \pT)^2}&= a_2+b_2\beta_2^2\;,\\\label{eq:3}
\lr{v_2^2\delta \pT} &= a_3-b_3\beta_2^3\cos(3\gamma)\;.
\end{align}
Here, $a_n$ and $b_n$ are positive coefficients that depend on the impact parameter. The $a_n$ terms capture contributions from quantum fluctuations, responsible for finite $v_2$ even at zero impact parameter~\cite{CMS:2013bza}, while the $b_n$ terms reflect the response to global deformation. Analogous relations can be written down for moments of initial conditions:
\begin{align}\nonumber
\lr{\varepsilon_2^2}&= a_1'+b_1'\beta_2^2\;,\\\nonumber
\lr{(\delta d_{\perp})^2}&= a_2'+b_2'\beta_2^2\;,\\\label{eq:3b}
\lr{\varepsilon_2^2 \delta d_{\perp}} &= a_3'-b_3'\beta_2^3\cos(3\gamma)\;.
\end{align}

Both $a_n$ and $b_n$ (and $a_n'$ and $b_n'$) are influenced by how colliding nucleons deposit energy in the overlap region. In phenomenological applications, the energy density is typically parametrized in a flexible way from $T_{\rm A,B}$, with those parameters inferred from experimental measurements. A commonly used ansatz is the generalized mean of the Trento model~\cite{Nijs:2023yab}:
\begin{align}\label{eq:4}
T(x,y) = \left(\frac{T_A^p+T_B^p}{2}\right)^{q/p}\;,
\end{align}
where $p$ and $q$ set the energy deposition prescription, usually with $q=1$. 

Consider the simplest scenario of head-on collisions involving spherical nuclei. In this case, the initial density distribution closely follows the nucleon density distribution in the $xy$-plane: $T(x,y)\approx T_{A} \approx T_{B}$. Introducing a deformation perturbation along an Euler angle, $T_A = T_{A,0}+ \delta_A(\Omega_A)$, one can show that the deformation effects are independent of parameter $p$, $T \approx T_0 + (\delta_A(\Omega_A)+\delta_B(\Omega_B))/2$. The spherical baseline $T_0(x,y)$ determines the $a_n$ coefficients. Meanwhile, the deformation-induced perturbations $\delta(x,y)$ vary with nuclear orientation and control the values of $b_n$. In this case, the initial geometry is determined mostly by the nucleon distribution~\footnote{However, in non-head-on collisions or when $q\neq1$, the linear dependence on nuclear deformation does not hold, and the relation between the initial energy distribution and nucleon distribution becomes non-trivial.}. Additionally, both $a_n$ and $b_n$ vary with $\sqrtsnn$. Gluon saturation effects at high energy tend to smooth out local fluctuations, reducing $a_n$. Conversely, the reduced central density could amplify the influence of nuclear deformation, increasing $b_n$.

\ul{\textit{Step3: Separating global deformation from quantum fluctuations}}.
The influence of nuclear deformation can be quite substantial. In head-on collisions of strongly-deformed uranium nuclei, the $b_n$ terms in Eq.~\ref{eq:3} can exceed 50\% of the baseline fluctuation contribution for $\lr{v_2^2}$ and $\lr{(\delta \pT)^2}$, and up to three times for $\lr{v_2^2\delta \pT}$~\cite{STAR:2024wgy}. While one could in principle constrain deformation by comparing hydrodynamic model calculations with data in a single collision system, both $a_n$ and $b_n$ are strongly influenced by final-state effects, limiting the precision of this approach.

A more robust strategy involves comparing two collision systems of similar mass but different shapes. Ratios of bulk observables from these isobaric systems minimize sensitivity to QGP transport properties, thereby exposing differences rooted in the initial conditions and nuclear shapes. The STAR Collaboration used this approach to determine deformation parameters for $^{238}$U by comparing collisions of $^{238}$U (strongly deformed) and $^{197}$Au (nearly spherical). For example, ratios of observables between a deformed nucleus and a spherical nucleus are: 
\begin{align}\nonumber
R_{\lr{v_2^2}}         &=1+\frac{b_{1}}{a_{1}}\beta_{\rm 2}^2 \;,\\\nonumber
R_{\lr{(\delta \pT)^2}} &=1+\frac{b_{2}}{a_{2}}\beta_{\rm 2}^2\;,\\\label{eq:5}
R_{\lr{v_2^2\delta \pT}}&=1-\frac{b_{3}}{a_{3}}\beta_{\rm 2}^3\cos(3\gamma)\;.
\end{align}
If the final-state responses for the $a_n$ and $b_n$ components are similar, the ratios $b_n/a_n$ primarily reflect the nuclear shape and, to a lesser extent, the energy deposition process.

\begin{figure}[htbp] \centering
\includegraphics[width=1\linewidth]{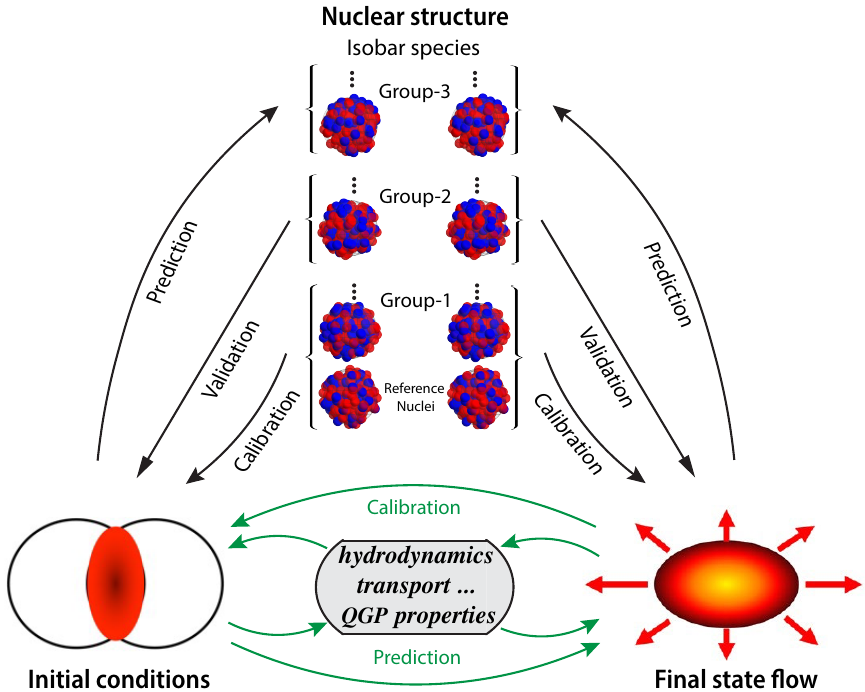}
\caption{\label{fig:5} {\bf Implementation of the imaging method.} 
The iterative processes of calibration, validation, and prediction to establish connections between nuclear structure and final-state flow observables or initial conditions structure, as well as the traditional approach of constraining initial conditions and QGP properties via hydrodynamic or transport model description of final state flow.}
\end{figure}

{\bf Implementation of the method. } To develop the imaging-by-smashing method toward practical application, we propose an iterative process by utilizing collisions of several isobaric or isobar-like nuclei (see Fig.~\ref{fig:5})
\begin{itemize}
\item {\bf Calibration.} Using species with known nuclear shapes to tune the response coefficients in Eqs.~\eqref{eq:3} and \eqref{eq:3b}. Ideally, one species should be nearly spherical, while others should have significant deformations to ensure accuracy in the extracted coefficients.
\item {\bf Validation.} Using additional species with known shapes to cross-check the response coefficients and expose potential limitations and inconsistencies.
\item {\bf Prediction.} Once the response coefficients are calibrated and validated, we may potentially make discoveries about unknown shapes from the measured flow observables in collisions of nuclei of interest. 
\end{itemize}
This iterative process aims to establish a direct connection between measured flow observables and nuclear shape in Eq.~\eqref{eq:3}, making this method fully data-driven. Within hydrodynamic or transport model frameworks, one can then systematically improve the implementation of initial conditions and explore the alternative definition of $\varepsilon_n$ and $d_{\perp}$ to achieve a consistent description of the experimental data. We are currently at the early calibration step, where reverse engineering using state-of-the-art hydrodynamic models shows encouraging agreement with shape parameters from low-energy methods~\cite{STAR:2024wgy,Zhang:2021kxj}. More collision species will enable the validation and prediction stages, making the method more data-driven. 

The reverse engineering idea holds at collision energies of a few tens of GeV or higher, where the distinct stages of the collision -- initial condition, QGP evolution, and freeze-out -- occur on well-separated timescales. At lower energies of below a few GeV, these timescales overlap, and the concept of initial conditions in terms of $\varepsilon_n$ and $d_{\perp}$ is no longer valid. Instead, nuclear structures themselves become the initial conditions of the collisions. Despite these complexities, Eq.~\eqref{eq:3} remains valid, except that the response coefficients $a_n$ and $b_n$ need to be calibrated using collisions of isobar species at the same $\sqrtsnn$. The ability to vary the nuclear shape, measure the flow response, and compare with model predictions remains a valuable approach for investigating the dynamics and properties of nuclear matter at lower energies. For instance, deformed nuclei offer a unique way to globally rearrange nucleons, allowing collisions of such nuclei to probe the nuclear equation of state. The Fermi momenta of nucleons are also relevant to the measured observables, leading to a more intricate interplay between nuclear structure and collision dynamics.


{\bf Imaging-by-smashing as a potential discovery tool.} 
The imaging-by-smashing technique offers exciting potential for advancing both heavy-ion and nuclear structure research~\cite{Jia:2022ozr}. However, realizing this potential requires rigorous calibration and thorough understanding of QGP initial conditions and evolution -- objectives already central to heavy-ion physics. A systematic study of isobaric or isobar-like nuclei with well-known structural properties shall provide crucial calibration benchmarks for validating this method.

It is important to clarify what this method measures. Rather than extracting intrinsic nuclear shapes via non-invasive methods as in low-energy nuclear physics, this technique extracts effective shape parameters that reflect how the underlying many-body wavefunction influences the outcomes of the dynamical collision process.  The nuclear structures accessed at high-energy heavy ion collisions are not entirely the same as those at low energies. While inherently model-dependent, I am hopeful that systematic validation across multiple collision systems will establish rigorous connections between nuclear structure and collision dynamics, making it a valuable complement to low-energy measurements.

Once properly calibrated and validated, imaging-by-smashing can become a powerful tool for exploring nuclear structure. The method is valuable for probing species whose shape parameters are difficult to determine through conventional techniques. The large number of particles produced per event enables exploration of higher-order deformations, including octupole and hexadecapole shapes, through higher-order flow harmonics. Multi-particle correlations can also be used to separate static deformation from dynamical shape fluctuations.

A particularly exciting application involves ``shape coexistence'', the phenomenon where some nuclei have multiple distinct shapes with nearly degenerate ground-state ($0^+$) energies~\cite{Garrett:2021kfb}. The nuclear ground-state wavefunction becomes an admixture of these different shape configurations. For instance, a superposition of two shapes can be represented as:
\begin{align}\nonumber
& \left|0_1^{+}\right\rangle=+\cos \eta\left|0_{\text {A }}^{+}\right\rangle+\sin \eta\left|0_{\text {B }}^{+}\right\rangle \\\label{eq:7}
& \left|0_2^{+}\right\rangle=-\sin \eta\left|0_{\text {A }}^{+}\right\rangle+\cos \eta\left|0_{\text {B }}^{+}\right\rangle\;,
\end{align}
where the mixing angle $\eta$ can be determined from electric quadrupole and monopole transitions between the two shape eigenstates. Some stable nuclei, such as $^{42}$Ca, $^{74}$Kr, and $^{152}$Sm, exhibit very strong mixing of two distinct shapes in their ground states. This represents a unique stable quantum superposition of collective many-body wavefunctions. Each high-energy collision should effectively collapse the nuclear wavefunction onto one specific shape component. The collision doesn't transform the nucleus into that shape; rather, it reveals the particular shape configuration present at the moment of impact. The presence of shape coexistence, depending on how this projection process is realized, could lead to different expectations in the measured flow fluctuations. High-energy collisions provide a unique setting to probe quantum superposition of collective wavefunctions, though this capability requires further theoretical and experimental development.

Imaging-by-smashing also offers new experimental insights into subbarrier nuclear fusion reactions, a quantum tunneling phenomenon critically influenced by the Coulomb barrier~\cite{Hagino:2012cu}. This barrier is sensitive to the orientation of colliding nuclei and the interplay between reaction timescale and timescales associated with nuclear deformations. Specifically, static nuclear deformations, characterized by longer rotational timescales, directly affect fusion probability, whereas dynamic shape fluctuations arising from nuclear vibrations, with their inherently shorter timescales, may have a smaller impact on the Coulomb barrier. The unique capability of high-energy collisions to differentiate between static deformation and dynamic shape fluctuations is therefore vital. This distinction may help decipher why sub-barrier fusion cross sections are significantly modified and determine whether static or dynamic nuclear properties exert a more important influence on this quantum tunneling process.

This method may also help address fundamental questions in nuclear physics. For instance, nuclear shape could generate nuclear electric dipole moment (EDM) through the laboratory Schiff moment that is proportional to the product of the quadrupole deformation parameter and square of the octupole deformation parameter $\beta_3^2\beta_2$~\cite{Spevak:1996tu}. One may use specially designed observables, such as correlation between triangular flow and radial flow, to constrain the Schiff moment, $\lr{v_3^2\delta[\pT]}\approx a-b\beta_3^2\beta_2$ (where $\beta_3^2$ arises from $v_3^2$, while $\beta_2$ arises from $\delta[\pT]$). $^{238}$U, though not a direct candidate for atomic EDM searches due to its even-even nature, could be used to characterize nuclear properties like its large static $\beta_2$ and a significant $\beta_3$ fluctuation~\cite{STAR:2025vbp} that are relevant to the Schiff moment, thus informing EDM measurements in odd-mass nuclei. The method may also be used to reduce theoretical uncertainties in neutrinoless double-beta decay searches ($0\nu\beta\beta$) \cite{Agostini:2022zub}, a process highly sensitive to the similarity of nuclear shapes in parent-daughter isobaric pairs. By comparing collisions of these nuclei, the technique can precisely quantify shape differences, thereby improving decay rate predictions. Furthermore, it opens new avenues for studying alpha clustering in light isobar-like systems such as $^{16}$O+$^{16}$O and $^{20}$Ne+$^{20}$Ne \cite{Giacalone:2024luz}, providing stringent tests for {\it ab initio} nuclear theories.

Various collider facilities offer further opportunities to develop this imaging method. For example, the NICA collider operates at lower center-of-mass energies ($\lesssim10$ GeV). It can collide a wide range of species, offering insights into how nuclear structure evolves across different energies and timescales. At RHIC, existing datasets from collisions such as $^{96}$Ru+$^{96}$Ru and $^{96}$Zr+$^{96}$Zr remain valuable for refining the imaging approach.

Planned system scans at the LHC beyond Run 3 (post-2029) promise to broaden the scope of this technique. A new ion source could allow up to four different collision species per heavy-ion run, increasing the range of nuclei available for study. Additionally, the SMOG-2 system at the LHCb experiment~\cite{2707819} allows the LHC ion beam to collide with fixed targets, offering greater flexibility in selecting collision systems. These advancements underscore the imaging-by-smashing method's potential to significantly advance our understanding of nuclear structure and heavy-ion physics across a broad range of energies.

{\bf Conclusion.} The imaging-by-smashing technique demonstrates the potential to turn the apparent destructiveness of high-energy collisions into a complementary imaging approach for nuclear structure studies. By exploiting the short timescales and collective dynamics of QGP formation and evolution, this method accesses nuclear many-body wavefunctions in ways that complement insights from conventional techniques. However, the successful application of this method requires major theoretical and experimental developments. Nevertheless, once properly validated, imaging-by-smashing could offer exciting possibilities, such as exploring shape coexistence, nuclear fusion, testing ab initio nuclear theories, and connecting nuclear structure to fundamental symmetries. I look forward to the successful development and validation of these techniques, and the eventual convergence of ultra-relativistic heavy-ion collisions with nuclear structure physics.
 
{\bf Acknowledgments.} The author would like to thank Aihong Tang for inviting him to present a BNL colloquium, which forms the basis of this note, as well as for valuable discussions with Peter Butler, Giuliano Giacalone, Kouichi Hagino, and John Woods. This work is supported by the U.S. Department of Energy, Award number DE-SC0024602.

\bibliography{../comment}{}

\begin{thebibliography}{27}%
\makeatletter
\providecommand \@ifxundefined [1]{%
 \@ifx{#1\undefined}
}%
\providecommand \@ifnum [1]{%
 \ifnum #1\expandafter \@firstoftwo
 \else \expandafter \@secondoftwo
 \fi
}%
\providecommand \@ifx [1]{%
 \ifx #1\expandafter \@firstoftwo
 \else \expandafter \@secondoftwo
 \fi
}%
\providecommand \natexlab [1]{#1}%
\providecommand \enquote  [1]{``#1''}%
\providecommand \bibnamefont  [1]{#1}%
\providecommand \bibfnamefont [1]{#1}%
\providecommand \citenamefont [1]{#1}%
\providecommand \href@noop [0]{\@secondoftwo}%
\providecommand \href [0]{\begingroup \@sanitize@url \@href}%
\providecommand \@href[1]{\@@startlink{#1}\@@href}%
\providecommand \@@href[1]{\endgroup#1\@@endlink}%
\providecommand \@sanitize@url [0]{\catcode `\\12\catcode `\$12\catcode
  `\&12\catcode `\#12\catcode `\^12\catcode `\_12\catcode `\%12\relax}%
\providecommand \@@startlink[1]{}%
\providecommand \@@endlink[0]{}%
\providecommand \url  [0]{\begingroup\@sanitize@url \@url }%
\providecommand \@url [1]{\endgroup\@href {#1}{\urlprefix }}%
\providecommand \urlprefix  [0]{URL }%
\providecommand \Eprint [0]{\href }%
\providecommand \doibase [0]{http://dx.doi.org/}%
\providecommand \selectlanguage [0]{\@gobble}%
\providecommand \bibinfo  [0]{\@secondoftwo}%
\providecommand \bibfield  [0]{\@secondoftwo}%
\providecommand \translation [1]{[#1]}%
\providecommand \BibitemOpen [0]{}%
\providecommand \bibitemStop [0]{}%
\providecommand \bibitemNoStop [0]{.\EOS\space}%
\providecommand \EOS [0]{\spacefactor3000\relax}%
\providecommand \BibitemShut  [1]{\csname bibitem#1\endcsname}%
\let\auto@bib@innerbib\@empty
\bibitem [{\citenamefont {{STAR Collaboration}}(2024)}]{STAR:2024wgy}%
  \BibitemOpen
  \bibfield  {author} {\bibinfo {author} {\bibnamefont {{STAR
  Collaboration}}},\ }\href {\doibase 10.1038/s41586-024-08097-2} {\bibfield
  {journal} {\bibinfo  {journal} {Nature}\ }\textbf {\bibinfo {volume} {635}},\
  \bibinfo {pages} {67} (\bibinfo {year} {2024})},\ \Eprint
  {http://arxiv.org/abs/2401.06625} {arXiv:2401.06625 [nucl-ex]} \BibitemShut
  {NoStop}%
\bibitem [{\citenamefont {Yang}\ \emph {et~al.}(2023)\citenamefont {Yang},
  \citenamefont {Wang}, \citenamefont {Wilkins},\ and\ \citenamefont
  {Garcia~Ruiz}}]{Yang:2022wbl}%
  \BibitemOpen
  \bibfield  {author} {\bibinfo {author} {\bibfnamefont {X.~F.}\ \bibnamefont
  {Yang}}, \bibinfo {author} {\bibfnamefont {S.~J.}\ \bibnamefont {Wang}},
  \bibinfo {author} {\bibfnamefont {S.~G.}\ \bibnamefont {Wilkins}}, \ and\
  \bibinfo {author} {\bibfnamefont {R.~F.}\ \bibnamefont {Garcia~Ruiz}},\
  }\href {\doibase 10.1016/j.ppnp.2022.104005} {\bibfield  {journal} {\bibinfo
  {journal} {Prog. Part. Nucl. Phys.}\ }\textbf {\bibinfo {volume} {129}},\
  \bibinfo {pages} {104005} (\bibinfo {year} {2023})}\BibitemShut {NoStop}%
\bibitem [{\citenamefont {Verney}(2025)}]{Verney:2025efj}%
  \BibitemOpen
  \bibfield  {author} {\bibinfo {author} {\bibfnamefont {D.}~\bibnamefont
  {Verney}},\ }\href {\doibase 10.1140/epja/s10050-025-01545-1} {\bibfield
  {journal} {\bibinfo  {journal} {Eur. Phys. J. A}\ }\textbf {\bibinfo {volume}
  {61}},\ \bibinfo {pages} {82} (\bibinfo {year} {2025})}\BibitemShut {NoStop}%
\bibitem [{\citenamefont {Busza}\ \emph {et~al.}(2018)\citenamefont {Busza},
  \citenamefont {Rajagopal},\ and\ \citenamefont {van~der
  Schee}}]{Busza:2018rrf}%
  \BibitemOpen
  \bibfield  {author} {\bibinfo {author} {\bibfnamefont {W.}~\bibnamefont
  {Busza}}, \bibinfo {author} {\bibfnamefont {K.}~\bibnamefont {Rajagopal}}, \
  and\ \bibinfo {author} {\bibfnamefont {W.}~\bibnamefont {van~der Schee}},\
  }\href {\doibase 10.1146/annurev-nucl-101917-020852} {\bibfield  {journal}
  {\bibinfo  {journal} {Ann. Rev. Nucl. Part. Sci.}\ }\textbf {\bibinfo
  {volume} {68}},\ \bibinfo {pages} {339} (\bibinfo {year} {2018})}\BibitemShut
  {NoStop}%
\bibitem [{\citenamefont {Yun}(2017)}]{Yun:2017}%
  \BibitemOpen
  \bibfield  {author} {\bibinfo {author} {\bibfnamefont {S.}~\bibnamefont
  {Yun}},\ }\href {\doibase 10.1038/s41598-017-18017-2} {\bibfield  {journal}
  {\bibinfo  {journal} {Scientific Reports}\ }\textbf {\bibinfo {volume} {7}}
  (\bibinfo {year} {2017}),\ 10.1038/s41598-017-18017-2}\BibitemShut {NoStop}%
\bibitem [{\citenamefont {O'Hara}\ \emph {et~al.}(2002)\citenamefont {O'Hara},
  \citenamefont {Hemmer}, \citenamefont {Gehm}, \citenamefont {Granade},\ and\
  \citenamefont {Thomas}}]{OHara:2002pqs}%
  \BibitemOpen
  \bibfield  {author} {\bibinfo {author} {\bibfnamefont {K.~M.}\ \bibnamefont
  {O'Hara}}, \bibinfo {author} {\bibfnamefont {S.~L.}\ \bibnamefont {Hemmer}},
  \bibinfo {author} {\bibfnamefont {M.~E.}\ \bibnamefont {Gehm}}, \bibinfo
  {author} {\bibfnamefont {S.~R.}\ \bibnamefont {Granade}}, \ and\ \bibinfo
  {author} {\bibfnamefont {J.~E.}\ \bibnamefont {Thomas}},\ }\href {\doibase
  10.1126/science.1079107} {\bibfield  {journal} {\bibinfo  {journal}
  {Science}\ }\textbf {\bibinfo {volume} {298}},\ \bibinfo {pages} {2179}
  (\bibinfo {year} {2002})},\ \Eprint {http://arxiv.org/abs/cond-mat/0212463}
  {arXiv:cond-mat/0212463} \BibitemShut {NoStop}%
\bibitem [{\citenamefont {Vager}\ \emph {et~al.}(1989)\citenamefont {Vager},
  \citenamefont {Naaman},\ and\ \citenamefont {Kanter}}]{cei}%
  \BibitemOpen
  \bibfield  {author} {\bibinfo {author} {\bibfnamefont {Z.}~\bibnamefont
  {Vager}}, \bibinfo {author} {\bibfnamefont {R.}~\bibnamefont {Naaman}}, \
  and\ \bibinfo {author} {\bibfnamefont {E.~P.}\ \bibnamefont {Kanter}},\
  }\href {\doibase 10.1126/science.244.4903.426} {\bibfield  {journal}
  {\bibinfo  {journal} {Science}\ }\textbf {\bibinfo {volume} {244}},\ \bibinfo
  {pages} {426} (\bibinfo {year} {1989})}\BibitemShut {NoStop}%
\bibitem [{\citenamefont {Boll}\ \emph {et~al.}(2022)\citenamefont {Boll},
  \citenamefont {Sch\"afer}, \citenamefont {Richard}, \citenamefont {Fehre},
  \citenamefont {Kastirke}, \citenamefont {Jurek},\ and\ \citenamefont
  {et~al.}}]{ceinature}%
  \BibitemOpen
  \bibfield  {author} {\bibinfo {author} {\bibfnamefont {R.}~\bibnamefont
  {Boll}}, \bibinfo {author} {\bibfnamefont {J.~M.}\ \bibnamefont {Sch\"afer}},
  \bibinfo {author} {\bibfnamefont {B.}~\bibnamefont {Richard}}, \bibinfo
  {author} {\bibfnamefont {K.}~\bibnamefont {Fehre}}, \bibinfo {author}
  {\bibfnamefont {G.}~\bibnamefont {Kastirke}}, \bibinfo {author}
  {\bibfnamefont {Z.}~\bibnamefont {Jurek}}, \ and\ \bibinfo {author}
  {\bibnamefont {et~al.}},\ }\href {\doibase
  doi.org/10.1038/s41567-022-01507-0} {\bibfield  {journal} {\bibinfo
  {journal} {Nature Physics}\ }\textbf {\bibinfo {volume} {18}},\ \bibinfo
  {pages} {423} (\bibinfo {year} {2022})}\BibitemShut {NoStop}%
\bibitem [{\citenamefont {{MADAI Collaboration}}()}]{madai}%
  \BibitemOpen
  \bibfield  {author} {\bibinfo {author} {\bibnamefont {{MADAI
  Collaboration}}},\ }\href {http://madai.phy.duke.edu/indexb842b842.html}
  {\enquote {\bibinfo {title} {{Models and Data Analysis Initiative}},}\
  }\BibitemShut {NoStop}%
\bibitem [{\citenamefont {Heinz}(2013)}]{Heinz:2013wva}%
  \BibitemOpen
  \bibfield  {author} {\bibinfo {author} {\bibfnamefont {U.~W.}\ \bibnamefont
  {Heinz}},\ }\href {\doibase 10.1088/1742-6596/455/1/012044} {\bibfield
  {journal} {\bibinfo  {journal} {J. Phys. Conf. Ser.}\ }\textbf {\bibinfo
  {volume} {455}},\ \bibinfo {pages} {012044} (\bibinfo {year} {2013})},\
  \Eprint {http://arxiv.org/abs/1304.3634} {arXiv:1304.3634 [nucl-th]}
  \BibitemShut {NoStop}%
\bibitem [{\citenamefont {Chatrchyan}\ \emph {et~al.}(2014)\citenamefont
  {Chatrchyan} \emph {et~al.}}]{CMS:2013bza}%
  \BibitemOpen
  \bibfield  {author} {\bibinfo {author} {\bibfnamefont {S.}~\bibnamefont
  {Chatrchyan}} \emph {et~al.} (\bibinfo {collaboration} {CMS}),\ }\href
  {\doibase 10.1007/JHEP02(2014)088} {\bibfield  {journal} {\bibinfo  {journal}
  {JHEP}\ }\textbf {\bibinfo {volume} {02}},\ \bibinfo {pages} {088} (\bibinfo
  {year} {2014})},\ \Eprint {http://arxiv.org/abs/1312.1845} {arXiv:1312.1845
  [nucl-ex]} \BibitemShut {NoStop}%
\bibitem [{\citenamefont {Jia}(2022{\natexlab{a}})}]{Jia:2021tzt}%
  \BibitemOpen
  \bibfield  {author} {\bibinfo {author} {\bibfnamefont {J.}~\bibnamefont
  {Jia}},\ }\href {\doibase 10.1103/PhysRevC.105.014905} {\bibfield  {journal}
  {\bibinfo  {journal} {Phys. Rev. C}\ }\textbf {\bibinfo {volume} {105}},\
  \bibinfo {pages} {014905} (\bibinfo {year} {2022}{\natexlab{a}})},\ \Eprint
  {http://arxiv.org/abs/2106.08768} {arXiv:2106.08768 [nucl-th]} \BibitemShut
  {NoStop}%
\bibitem [{\citenamefont {Niemi}\ \emph {et~al.}(2013)\citenamefont {Niemi},
  \citenamefont {Denicol}, \citenamefont {Holopainen},\ and\ \citenamefont
  {Huovinen}}]{Niemi:2012aj}%
  \BibitemOpen
  \bibfield  {author} {\bibinfo {author} {\bibfnamefont {H.}~\bibnamefont
  {Niemi}}, \bibinfo {author} {\bibfnamefont {G.~S.}\ \bibnamefont {Denicol}},
  \bibinfo {author} {\bibfnamefont {H.}~\bibnamefont {Holopainen}}, \ and\
  \bibinfo {author} {\bibfnamefont {P.}~\bibnamefont {Huovinen}},\ }\href
  {\doibase 10.1103/PhysRevC.87.054901} {\bibfield  {journal} {\bibinfo
  {journal} {Phys. Rev. C}\ }\textbf {\bibinfo {volume} {87}},\ \bibinfo
  {pages} {054901} (\bibinfo {year} {2013})}\BibitemShut {NoStop}%
\bibitem [{\citenamefont {Bo\'zek}\ and\ \citenamefont
  {Broniowski}(2012)}]{Bozek:2012fw}%
  \BibitemOpen
  \bibfield  {author} {\bibinfo {author} {\bibfnamefont {P.}~\bibnamefont
  {Bo\'zek}}\ and\ \bibinfo {author} {\bibfnamefont {W.}~\bibnamefont
  {Broniowski}},\ }\href {\doibase 10.1103/PhysRevC.85.044910} {\bibfield
  {journal} {\bibinfo  {journal} {Phys. Rev. C}\ }\textbf {\bibinfo {volume}
  {85}},\ \bibinfo {pages} {044910} (\bibinfo {year} {2012})}\BibitemShut
  {NoStop}%
\bibitem [{\citenamefont {Bernhard}\ \emph {et~al.}(2019)\citenamefont
  {Bernhard}, \citenamefont {Moreland},\ and\ \citenamefont
  {Bass}}]{Bernhard:2019bmu}%
  \BibitemOpen
  \bibfield  {author} {\bibinfo {author} {\bibfnamefont {J.~E.}\ \bibnamefont
  {Bernhard}}, \bibinfo {author} {\bibfnamefont {J.~S.}\ \bibnamefont
  {Moreland}}, \ and\ \bibinfo {author} {\bibfnamefont {S.~A.}\ \bibnamefont
  {Bass}},\ }\href {\doibase 10.1038/s41567-019-0611-8} {\bibfield  {journal}
  {\bibinfo  {journal} {Nature Phys.}\ }\textbf {\bibinfo {volume} {15}},\
  \bibinfo {pages} {1113} (\bibinfo {year} {2019})}\BibitemShut {NoStop}%
\bibitem [{\citenamefont {Jia}\ \emph {et~al.}(2024)\citenamefont {Jia} \emph
  {et~al.}}]{Jia:2022ozr}%
  \BibitemOpen
  \bibfield  {author} {\bibinfo {author} {\bibfnamefont {J.}~\bibnamefont
  {Jia}} \emph {et~al.},\ }\href {\doibase 10.1007/s41365-024-01589-w}
  {\bibfield  {journal} {\bibinfo  {journal} {Nucl. Sci. Tech.}\ }\textbf
  {\bibinfo {volume} {35}},\ \bibinfo {pages} {220} (\bibinfo {year} {2024})},\
  \Eprint {http://arxiv.org/abs/2209.11042} {arXiv:2209.11042 [nucl-ex]}
  \BibitemShut {NoStop}%
\bibitem [{\citenamefont {Giacalone}(2020)}]{Giacalone:2019pca}%
  \BibitemOpen
  \bibfield  {author} {\bibinfo {author} {\bibfnamefont {G.}~\bibnamefont
  {Giacalone}},\ }\href {\doibase 10.1103/PhysRevLett.124.202301} {\bibfield
  {journal} {\bibinfo  {journal} {Phys. Rev. Lett.}\ }\textbf {\bibinfo
  {volume} {124}},\ \bibinfo {pages} {202301} (\bibinfo {year} {2020})},\
  \Eprint {http://arxiv.org/abs/1910.04673} {arXiv:1910.04673 [nucl-th]}
  \BibitemShut {NoStop}%
\bibitem [{\citenamefont {Jia}(2022{\natexlab{b}})}]{Jia:2021qyu}%
  \BibitemOpen
  \bibfield  {author} {\bibinfo {author} {\bibfnamefont {J.}~\bibnamefont
  {Jia}},\ }\href {\doibase 10.1103/PhysRevC.105.044905} {\bibfield  {journal}
  {\bibinfo  {journal} {Phys. Rev. C}\ }\textbf {\bibinfo {volume} {105}},\
  \bibinfo {pages} {044905} (\bibinfo {year} {2022}{\natexlab{b}})},\ \Eprint
  {http://arxiv.org/abs/2109.00604} {arXiv:2109.00604 [nucl-th]} \BibitemShut
  {NoStop}%
\bibitem [{\citenamefont {Nijs}\ and\ \citenamefont {van~der
  Schee}(2023)}]{Nijs:2023yab}%
  \BibitemOpen
  \bibfield  {author} {\bibinfo {author} {\bibfnamefont {G.}~\bibnamefont
  {Nijs}}\ and\ \bibinfo {author} {\bibfnamefont {W.}~\bibnamefont {van~der
  Schee}},\ }\href@noop {} {\  (\bibinfo {year} {2023})},\ \Eprint
  {http://arxiv.org/abs/2304.06191} {arXiv:2304.06191 [nucl-th]} \BibitemShut
  {NoStop}%
\bibitem [{\citenamefont {Zhang}\ and\ \citenamefont
  {Jia}(2022)}]{Zhang:2021kxj}%
  \BibitemOpen
  \bibfield  {author} {\bibinfo {author} {\bibfnamefont {C.}~\bibnamefont
  {Zhang}}\ and\ \bibinfo {author} {\bibfnamefont {J.}~\bibnamefont {Jia}},\
  }\href {\doibase 10.1103/PhysRevLett.128.022301} {\bibfield  {journal}
  {\bibinfo  {journal} {Phys. Rev. Lett.}\ }\textbf {\bibinfo {volume} {128}},\
  \bibinfo {pages} {022301} (\bibinfo {year} {2022})},\ \Eprint
  {http://arxiv.org/abs/2109.01631} {arXiv:2109.01631 [nucl-th]} \BibitemShut
  {NoStop}%
\bibitem [{\citenamefont {Garrett}\ \emph {et~al.}(2022)\citenamefont
  {Garrett}, \citenamefont {Zieli\'nska},\ and\ \citenamefont
  {Cl\'ement}}]{Garrett:2021kfb}%
  \BibitemOpen
  \bibfield  {author} {\bibinfo {author} {\bibfnamefont {P.~E.}\ \bibnamefont
  {Garrett}}, \bibinfo {author} {\bibfnamefont {M.}~\bibnamefont
  {Zieli\'nska}}, \ and\ \bibinfo {author} {\bibfnamefont {E.}~\bibnamefont
  {Cl\'ement}},\ }\href {\doibase 10.1016/j.ppnp.2021.103931} {\bibfield
  {journal} {\bibinfo  {journal} {Prog. Part. Nucl. Phys.}\ }\textbf {\bibinfo
  {volume} {124}},\ \bibinfo {pages} {103931} (\bibinfo {year}
  {2022})}\BibitemShut {NoStop}%
\bibitem [{\citenamefont {Hagino}\ and\ \citenamefont
  {Takigawa}(2012)}]{Hagino:2012cu}%
  \BibitemOpen
  \bibfield  {author} {\bibinfo {author} {\bibfnamefont {K.}~\bibnamefont
  {Hagino}}\ and\ \bibinfo {author} {\bibfnamefont {N.}~\bibnamefont
  {Takigawa}},\ }\href {\doibase 10.1143/PTP.128.1061} {\bibfield  {journal}
  {\bibinfo  {journal} {Prog. Theor. Phys.}\ }\textbf {\bibinfo {volume}
  {128}},\ \bibinfo {pages} {1061} (\bibinfo {year} {2012})},\ \Eprint
  {http://arxiv.org/abs/1209.6435} {arXiv:1209.6435 [nucl-th]} \BibitemShut
  {NoStop}%
\bibitem [{\citenamefont {Spevak}\ \emph {et~al.}(1997)\citenamefont {Spevak},
  \citenamefont {Auerbach},\ and\ \citenamefont {Flambaum}}]{Spevak:1996tu}%
  \BibitemOpen
  \bibfield  {author} {\bibinfo {author} {\bibfnamefont {V.}~\bibnamefont
  {Spevak}}, \bibinfo {author} {\bibfnamefont {N.}~\bibnamefont {Auerbach}}, \
  and\ \bibinfo {author} {\bibfnamefont {V.~V.}\ \bibnamefont {Flambaum}},\
  }\href {\doibase 10.1103/PhysRevC.56.1357} {\bibfield  {journal} {\bibinfo
  {journal} {Phys. Rev. C}\ }\textbf {\bibinfo {volume} {56}},\ \bibinfo
  {pages} {1357} (\bibinfo {year} {1997})},\ \Eprint
  {http://arxiv.org/abs/nucl-th/9612044} {arXiv:nucl-th/9612044} \BibitemShut
  {NoStop}%
\bibitem [{\citenamefont {{STAR Collaboration}}(2025)}]{STAR:2025vbp}%
  \BibitemOpen
  \bibfield  {author} {\bibinfo {author} {\bibnamefont {{STAR
  Collaboration}}},\ }\href@noop {} {\  (\bibinfo {year} {2025})},\ \Eprint
  {http://arxiv.org/abs/2506.17785} {arXiv:2506.17785 [nucl-ex]} \BibitemShut
  {NoStop}%
\bibitem [{\citenamefont {Agostini}\ \emph {et~al.}(2023)\citenamefont
  {Agostini}, \citenamefont {Benato}, \citenamefont {Detwiler}, \citenamefont
  {Men\'endez},\ and\ \citenamefont {Vissani}}]{Agostini:2022zub}%
  \BibitemOpen
  \bibfield  {author} {\bibinfo {author} {\bibfnamefont {M.}~\bibnamefont
  {Agostini}}, \bibinfo {author} {\bibfnamefont {G.}~\bibnamefont {Benato}},
  \bibinfo {author} {\bibfnamefont {J.~A.}\ \bibnamefont {Detwiler}}, \bibinfo
  {author} {\bibfnamefont {J.}~\bibnamefont {Men\'endez}}, \ and\ \bibinfo
  {author} {\bibfnamefont {F.}~\bibnamefont {Vissani}},\ }\href {\doibase
  10.1103/RevModPhys.95.025002} {\bibfield  {journal} {\bibinfo  {journal}
  {Rev. Mod. Phys.}\ }\textbf {\bibinfo {volume} {95}},\ \bibinfo {pages}
  {025002} (\bibinfo {year} {2023})},\ \Eprint
  {http://arxiv.org/abs/2202.01787} {arXiv:2202.01787 [hep-ex]} \BibitemShut
  {NoStop}%
\bibitem [{\citenamefont {Giacalone}\ \emph {et~al.}(2025)\citenamefont
  {Giacalone} \emph {et~al.}}]{Giacalone:2024luz}%
  \BibitemOpen
  \bibfield  {author} {\bibinfo {author} {\bibfnamefont {G.}~\bibnamefont
  {Giacalone}} \emph {et~al.},\ }\href {\doibase 10.1103/k8rb-jgvq} {\bibfield
  {journal} {\bibinfo  {journal} {Phys. Rev. Lett.}\ }\textbf {\bibinfo
  {volume} {135}},\ \bibinfo {pages} {012302} (\bibinfo {year} {2025})},\
  \Eprint {http://arxiv.org/abs/2402.05995} {arXiv:2402.05995 [nucl-th]}
  \BibitemShut {NoStop}%
\bibitem [{\citenamefont {{LHCb Collaboration}}()}]{2707819}%
  \BibitemOpen
  \bibfield  {author} {\bibinfo {author} {\bibnamefont {{LHCb
  Collaboration}}},\ }\href {\doibase 10.17181/CERN.SAQC.EOWH} {\
  10.17181/CERN.SAQC.EOWH}\BibitemShut {NoStop}%
\end{thebibliography}%

\end{document}